\documentclass{aa} 

\usepackage{graphicx}
\usepackage{txfonts}
\usepackage{color}

\begin{document}

   \title{Full interferometric map of the L1157 southern outflow:\\ Formamide (NH$_2$CHO) can form in the gas, after all}
   \titlerunning{Full formamide map of the L1157 southern outflow}


   \author{A. L\'opez-Sepulcre
          \inst{1,2}
          \and
          C. Codella\inst{3,1}
          \and
          C. Ceccarelli\inst{1}
          \and
          L. Podio\inst{3}
          \and
          J. Robuschi\inst{1}
          \fnmsep\thanks{This work is based on observations carried out under project number I17AB with the IRAM NOEMA Interferometer}
          }

   \institute{Univ. Grenoble Alpes, CNRS, IPAG, 38000 Grenoble, France\\
              \email{ana.lopez-sepulcre@univ-grenoble-alpes.fr}
         \and
Institut de Radioastronomie Millimétrique, 300 rue de la Piscine, Domaine Universitaire, 38406 Saint-Martin d’Hères, France
         \and
             INAF, Osservatorio Astrofisico di Arcetri, Largo E. Fermi 5, 50125 Firenze, Italy
             }

   \date{Received ; accepted }

 
  \abstract
   {The formation mechanism of interstellar formamide (NH$_2$CHO), a key prebiotic precursor, is currently a matter of hot debate within the astrochemistry community, with both gas-phase and grain-surface chemical pathways having been proposed as its dominant formation route.}
   {The aim of the present study is to place firm observational constraints on the formation pathways leading to formamide thanks to new interferometric observations of the molecular outflow driven by the protostellar binary L1157.}
   {We employed the IRAM NOEMA interferometer to map the entire southern outflow of L1157, which contains three main shocked regions with increasing post-shock age: B0, B1, and B2. This allowed us to measure how the abundance of formamide, that of acetaldehyde (CH$_3$CHO), and the ratio of the two, vary with time in this region. In order to gain a greater understanding of the most likely formation routes of formamide, we ran a grid of astrochemical models and compared these to our observations.}
   {A comparison between observations and astrochemical modelling indicates that there are two possible scenarios: one in which the amount of formamide observed can be explained by gas-phase-only chemistry, and more specifically via the reaction H$_2$CO + NH$_2$ $\rightarrow$ NH$_2$CHO + H$_2$, and another in which part of the observed formamide originates from surface chemistry and part from gas-phase chemistry. Surface chemistry alone cannot account for the abundance of formamide that we measure.}
   {While grain-surface chemistry cannot be ruled out, the present study brings definitive proof that gas-phase chemistry does work in L1157-B and acts efficiently in the production of this molecular species.}


   \maketitle
%

\section{Introduction}\label{intro}

Life represents the highest level of chemical complexity that we know of today, and yet there is a surprising simplicity to it, in that all living beings are made of the same basic ingredients: organic molecules. Many of the simplest versions of these organic molecules can be detected in the interstellar medium (ISM), in particular in sites of star and planet formation. Within the family of interstellar organic molecules detected to date, there is one that stands out due to its prebiotic interest: formamide (NH$_2$CHO). Indeed, formamide was identified by biochemists as a key prebiotic precursor whose chemical versatility can lead, with the right catalists, to a great variety of larger molecules and compounds essential to life, including proteins and nucleic bases (e.g. \citealp{saladino12,botta18,ferus22}). This, coupled with its detection in a variety of interstellar environments, including hot corinos, high-mass star forming regions, and even external galaxies (see e.g. \citealp{ls19} and references therein), has awakened considerable interest in the astrochemistry community, and numerous efforts have been devoted in the past decade or so to understanding how it is formed (and destroyed) in such environments. Today, its formation mechanisms are still hotly debated, with both gas-phase \citep{barone15,skouteris17} and grain-surface chemistry pathways \citep{rimola18,Enrique-Romero2022} being plausible possibilities (see Sect.\,\ref{subsec:astro-model-chemistry} for a detailed discussion on interstellar formamide chemistry).

The aim of the present study is to shed more light on this debate by targeting a series of protostellar shocked
regions lying along the outflow driven by L1157  with the IRAM (Institut de Radio Astronomie Millim\'etrique) NOEMA (NOrthern Extended Millimeter Array) interferometer. These are excellent targets with which to explore the time-dependent gas-phase chemistry of molecules, and formamide in particular.

The protostellar binary system associated with L1157-mm \citep{tobin22} drives an episodic and precessing jet \citep{gueth96,gueth98,podio16}, which, in turn, opens several outflow cavities \citep{gueth96}. In particular, the southern blueshifted outflow is associated with three main shocked regions produced by jet or cavity impacts, called B0, B1, and B2. The southern lobe is a well-known laboratory with which to study  the alteration of gas chemistry by the injection of species that were previously frozen onto the dust mantles or were sputtered directly from the refractory core.
Moreover, the gas-phase abundances of a large number of molecules (e.g. CH$_3$OH, H$_2$CO, SiO, and S-bearing species), as observed with single dishes and interferometers, are significantly enhanced in the lobe, making L1157 the archetype of the so-called chemically rich outflows \citep{bachiller01,benedettini07,Yamaguchi2012}.
Several interstellar complex organic molecules (iCOMs), that is, carbon-based molecules with at least six atoms that typically contain other heavy elements such as nitrogen or oxygen, have also been detected: first methanol (CH$_3$OH; \citealp{Bachiller1995,Bachiller1997}), and successively methyl formate (HCOOCH$_3$), formic acid (HCOOH), methyl cyanide (CH$_3$CN), ethanol (C$_2$H$_5$OH), and dimethyl ether (CH$_3$OCH$_3$; \citealp{arce08,lefloch17}).

The brightest cavity as seen in CO emission, B1, consists of a series of shocked spots caused by different episodes of ejection impacting against the cavity wall, and has been mapped using the IRAM Plateau de Bure/NOEMA array in CH$_3$OH \citep{benedettini07,clau20}, CH$_3$CN \citep{clau09}, NH$_2$CHO \citep{clau17}, and CH$_3$CHO \citep{clau15,clau20}. These images solidified the status of the shocked spots in L1157-B1 as an interstellar laboratory for studies of shock-driven chemical complexity. Among them, the work carried out by \cite{clau17} is of particular relevance here, where the authors mapped and analysed the spatial distribution of NH$_2$CHO and CH$_3$CHO emission in L1157-B1 as part of the NOEMA Large Program SOLIS (Seeds Of Life In Space; \citealp{cec17}). These authors noted that NH$_2$CHO is most intense and abundant in the older shocked spots of the region, while CH$_3$CHO is more prominent in the younger ones. Their astrochemical modelling can only explain these findings if gas-phase chemistry is the main actor in the synthesis of NH$_2$CHO. While their study placed valuable constraints on the chemical pathways leading to formamide, it was limited to a relatively small region of the entire outflow lobe, where the evolution of chemistry cannot be probed on a sufficiently large time range to consolidate their results.

In the present study, we follow up on the above-mentioned work by mapping the emission of NH$_2$CHO and CH$_3$CHO along the full L1157 southern lobe, from the protobinary position down to the older B2 shocked region. Our goal is to measure the changes in column density of these two molecular species throughout the southern outflow lobe in order to evaluate the time-dependent chemical evolution of their relative abundances, and place constraints on their formation and destruction pathways. Our results provide more solid grounds for the confirmation or revision of the results of \cite{clau17}, which favour gas-phase chemistry as the dominant route to form NH$_2$CHO.

\section{Observations}\label{obs}

The IRAM NOEMA interferometer array was employed during several runs between 22 December 2017 and 11 May 2018 in its nine-element D (i.e. compact) configuration to build an eight-field mosaic of the southern outflow driven by the L1157 protobinary, covering a total map area of 2.5\,arcmin$^2$. The centre coordinates of the mosaic are Right Ascension (RA)(J2000)\,$=$\,20$^\mathrm{h}$39$^\mathrm{m}$09.635$^\mathrm{s}$, Declination (Dec.)(J2000)\,$=$\,68$^\circ$01$'$19.80$''$. The PolyFiX correlator was used to image several molecular lines of NH$_2$CHO and CH$_3$CHO in the 3\,mm frequency band (see Table\,\ref{tlines}). For this purpose, high-spectral-resolution windows ---each with a nominal channel width of 62.5\,kHz--- were placed at the frequencies of the targeted molecular transitions.

\begin{table*}[!thb]
\centering
\caption{Observed molecular transitions and their corresponding map properties}
\begin{tabular}{llccccccc}
\hline
Molecule & Transition & $\nu_\mathrm{rest}$\tablefootmark{a} & g$_\mathrm{up}$\tablefootmark{a} & E$_\mathrm{up}$\tablefootmark{a} & A$_{ij}$\tablefootmark{a} & Beam size & Beam P.A. & 1$\sigma$ RMS\tablefootmark{b}\\
 & & (MHz) & & (K) & ($10^{-5}$\,s$^{-1}$) & ($'' \times ''$) & (deg) & mJy\,beam$^{-1}$ \\
\hline
\hline
NH$_2$CHO & 4$_{1,4}-3_{1,3}$ & 81\,693.446 & 9 & 12.8 & 3.46 & $4.9 \times 4.5$ & $-42.9$ & 5.0 \\
NH$_2$CHO & 4$_{0,4}-3_{0,3}$ & 84\,542.330 & 9 & 10.2 & 4.09 & $4.8 \times 4.4$ & $137$ & 4.6 \\
NH$_2$CHO & 4$_{1,3}-3_{1,2}$ & 87\,848.874 & 9 & 13.5 & 4.30 & $4.6 \times 4.2$ & $-43.5$ & 4.5 \\
\hline
CH$_3$CHO & 5$_{1,4}-4_{1,3}$\,E & 98\,863.314 & 22 & 16.6 & 3.10 & $5.5 \times 3.3$ & $99.4$ & 10.5 \\
CH$_3$CHO & 5$_{1,4}-4_{1,3}$\,A & 98\,900.945 & 22 & 16.5 & 3.11 & $5.5 \times 3.3$ & $99.4$ & 9.9 \\
\hline
CS & 2--1 & 97\,980.953 & 5 & 7.1 & 1.67 & $4.2 \times 3.9$ & $-43.8$ & 1.0\\
\hline
\end{tabular}
\tablefoot{
\tablefoottext{a}{Extracted from the Cologne Database for Molecular Spectroscopy (CDMS; \citealp{muller05,endres16}) for NH$_2$CHO and CS, and from the Jet Propulsion Laboratory (JPL) database for CH$_3$CHO.}
\tablefoottext{b}{Measured in a 0.5\,km\,s$^{-1}$ channel, except for CS, whose cube has a channel width of 2\,MHz (7\,km\,s$^{-1}$).}
}
\label{tlines}
\end{table*}

The quasars $2010+723$, $1928+738$, and J$1933+656$ were used as phase calibrators, and the flux scale was calibrated by observing MWC349 and LKHA101. The estimated uncertainty on the calibrated absolute flux densities is below 10\%.

The data were calibrated using the CLIC software of the GILDAS package. Imaging and deconvolution were performed with the GILDAS MAPPING software, with natural weighting to maximise sensitivity. Subtraction of the continuum emission arising entirely from the protostar L1157-mm was carried out in the visibility plane. To improve the signal-to-noise ratio (S/N), the spectral resolution of the cubes was smoothed to a channel width of 0.5\,km\,s$^{-1}$. The resulting beam size and 1$\sigma$ RMS achieved for each detected molecular line are listed in Table\,\ref{tlines}.


\section{Results}\label{results}

We detected three lines of NH$_2$CHO and two lines of CH$_3$CHO  with $S/N>5$. The upper state energies of the associated transitions, $E_{u}$, range between 10.2\,K and 16.6\,K, as indicated in Table\,\ref{tlines}. So far, multiple detailed studies of iCOMs and other molecular tracers have been carried out towards the B1 shocked region (e.g. \citealp{fontani14,clau15,clau17,clau20,lefloch17,spezzano20}). However, and excepting the CH$_3$OH maps obtained by \cite{benedettini07}, this is the first time we can investigate iCOMs such as formamide and acetaldehyde along the entire southern outflow lobe. This allows us to identify time-dependent trends that constrain the formation routes of the analysed molecules with greater empirical support than ever before.

\subsection{Maps and spectra}

\begin{figure*}[!bht]
   \centering
   \includegraphics[scale=0.7]{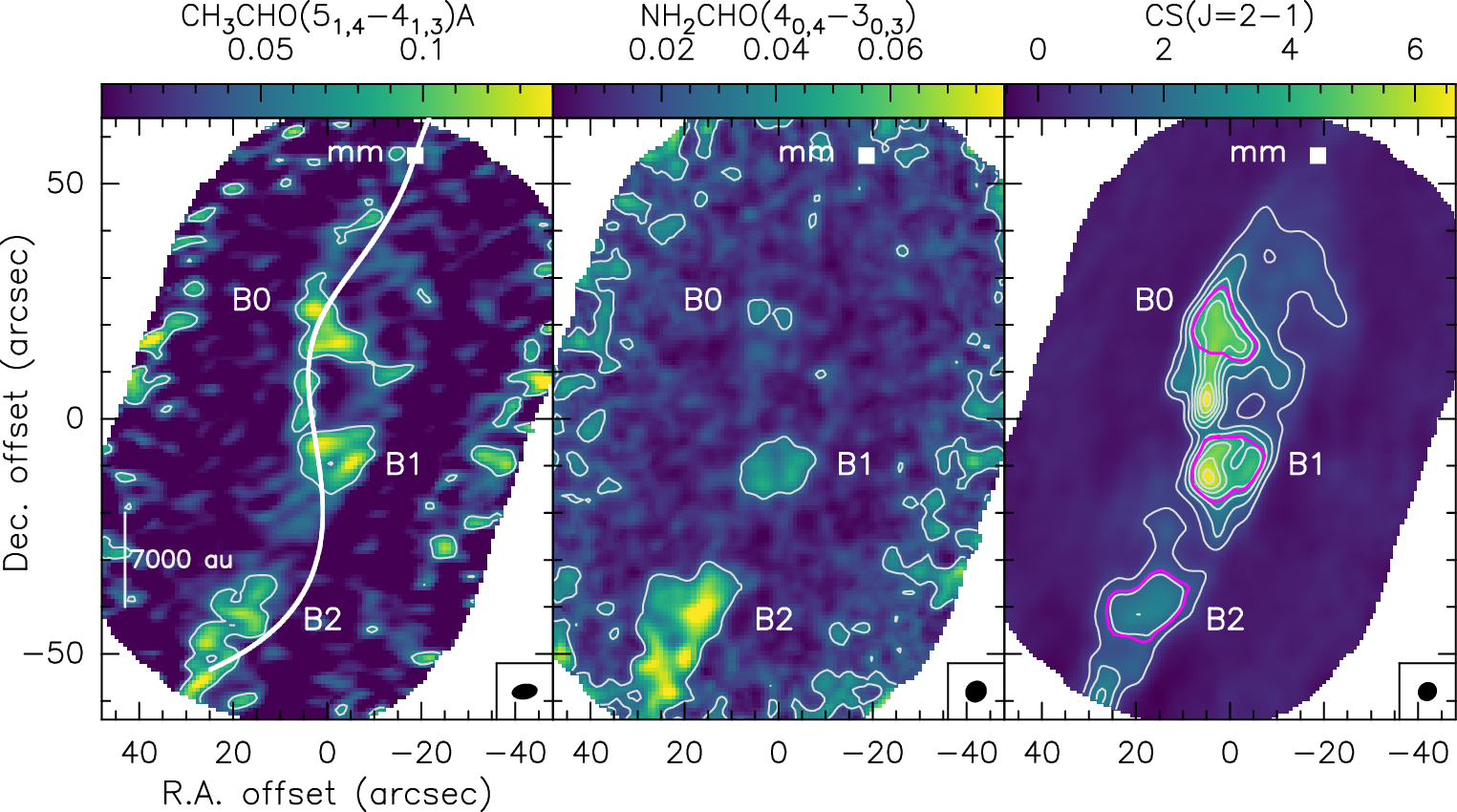}
   \caption{Velocity-integrated line maps of CH$_3$CHO(5$_{1,4}-4_{1,3}$)A (\textit{left}), NH$_2$CHO(4$_{0,4}-3_{0,3}$) (\textit{centre}), and CS(J=2-1) (\textit{right}) along the L1157 southern outflow. The colour scale is displayed at the top of each panel in units of Jy\,beam$^{-1}$ km\,s$^{-1}$. For CH$_3$CHO and NH$_2$CHO, the 4$\sigma$ contours are displayed in grey, with 1$\sigma$ values being 15 and 7\,mJy\,beam$^{-1}$\,km\,s$^{-1}$, respectively. For CS, grey contours start at 50$\sigma$ and increase in steps of 100$\sigma$, with 1$\sigma = 18$\,mJy\,beam$^{-1}$\,km\,s$^{-1}$. This map has been used to define the three polygons B0, B1, and B2 (magenta), which are used to derive column densities. The (0$''$,0$''$) position corresponds to RA(J2000)\,$=$\,20$^\mathrm{h}$39$^\mathrm{m}$09.635$^\mathrm{s}$, Dec.(J2000)\,$=$\,68$^\circ$01$'$19.80$''$. The synthesised beams are depicted in the lower-right corner of each panel. The precession model computed by \citet{podio16} is marked in the left panel (white line). The white squares at the top of each panel correspond to the position of L1157-mm, the protobinary driving the outflow \citep{bachiller01,lefloch17}.}
              \label{fmaps}%
    \end{figure*}

Figure\,\ref{fmaps} shows the velocity-integrated maps of the two targeted molecules, where we have selected the most intense emission line of each. We note that the maps of the other detected molecular lines display very similar emission distributions per species. The velocity range of integration for these maps is ($-2.5$,4.5)\,km\,s$^{-1}$, which corresponds to the range where NH$_2$CHO emission is detected above 2$\sigma$. We estimated the flux recovered by the interferometer by comparing it with the line fluxes measured on B1 with the ASAI (Astrochemical Surveys At IRAM, \citealp{lefloch18}) spectra obtained with the IRAM 30 m telescope. We find that we recover 84\% of the flux for both molecular lines (see Fig\,\ref{floss}). Assuming a similar amount of filtered flux for the other regions along the outflow (B0 and B2), we can safely compare molecular tracers between them, knowing that they are only slightly affected by interferometric flux filtering. 

As can be seen in the maps, we detect the two targeted species along the entire blueshifted outflow lobe. The three main shocked regions, B0, B1, and B2, are all detected with varying relative degrees of intensity. Formamide emits most intensely towards the south, particularly in B2. On the other hand, acetaldehyde has slightly more intense emission in B0 and B1 relative to B2. As expected in mosaics, the edges of each map are noisier than the more central areas, where the noise is lower and more homogeneous.

\begin{figure}
   \centering
   \includegraphics[scale=0.65]{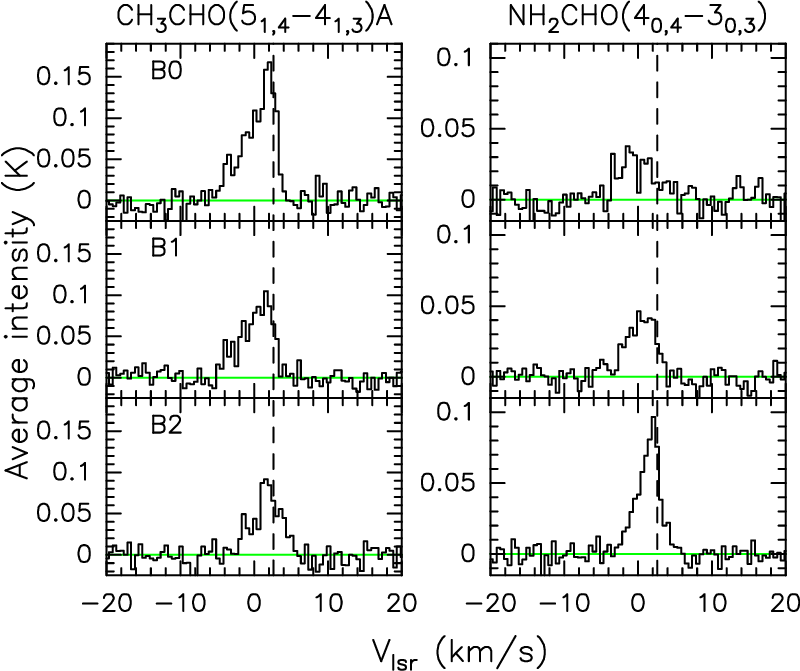}
   \caption{Spectra of CH$_3$CHO(5$_{1,4}-4_{1,3}$)A and NH$_2$CHO(4$_{0,4}-3_{0,3}$) averaged over the three polygons defined in Fig.\,\ref{fmaps} for the shocked regions B0, B1, and B2. The dashed vertical line and the green horizontal line mark, respectively, the systemic velocity of L1157-mm ($V_\mathrm{lsr}=2.6$\,km\,s$^{-1}$) and the zero-intensity level of each spectrum.}
              \label{fspt}%
    \end{figure}

A third map is presented in Fig.\,\ref{fmaps} displaying the bright CS(2--1) velocity-integrated emission, which in this work is used solely as a reference to define the polygons in B0, B1, and B2, which are used to derive the corresponding NH$_2$CHO and CH$_3$CHO column densities (see Sect.\,\ref{ratio}). These polygons have been defined following the 180$\sigma$ CS(2--1) contour in the case of B0 and B1, and the $90\sigma$ contour in the case of B2. As shown in the right panel of Fig.\,\ref{fmaps}, the B0 polygon has been truncated to include only the northernmost, presumably younger emission. The line intensity trends observed in Fig.\,\ref{fmaps} are also evident when examining the spectra averaged over these three polygons (Fig.\,\ref{fspt}). As expected in the southern outflow lobe, the lines are blueshifted with respect to the protostellar systemic velocity, $V_\mathrm{lsr}=2.6$\,km\,s$^{-1}$. It is clear that NH$_2$CHO and CH$_3$CHO have opposite intensity trends when moving from the northern B0 position to the southern B2 position.

\subsection{Column density ratios}\label{ratio}

Given the unavailability of collisional coefficients and the limited range of $E_{u}$ values covered by the two molecules, we derived molecular column densities adopting a range of excitation temperatures of $T_\mathrm{ex}=\,$10--20\,K, assuming local thermodynamical equilibrium (LTE) conditions and optically thin line emission. Previous studies have shown that LTE provides molecular column densities consistent with those derived using non-LTE methods within a factor of about 2. Indeed, \cite{clau20} found this to be true for methanol (CH$_3$OH) in L1157-B1 using both LTE and non-LTE methods. \cite{lefloch12} also concluded using CO line emission in L1157-B1 that the excitation conditions in this region are close to LTE. Regarding the assumption of optically thin lines, we verified {a posteriori} that they are indeed optically thin ($\tau<0.01$) for the derived column density values. The chosen range of $T_\mathrm{ex}$ is based on previous studies where rotational diagrams were obtained for both NH$_2$CHO and CH$_3$CHO in B1 \citep{clau20,mendoza14} and for NH$_2$CHO in B2 \citep{mendoza14}. In the case of formamide, we were able to put quantitative constraints on the rotational temperature, $T_\mathrm{rot}$, using 3$\sigma$ upper limits of several undetected transitions in our dataset (see Appendix\,\ref{appb} for details). In short, we find that it is lower than 20\, and 30\,K in B2 and B1, respectively, while it is higher than 10\,K in B0. Our adopted range of excitation temperatures therefore appears to be reasonable, although for B0 there is a larger uncertainty.

\begin{figure}
   \centering
   \includegraphics[scale=0.68]{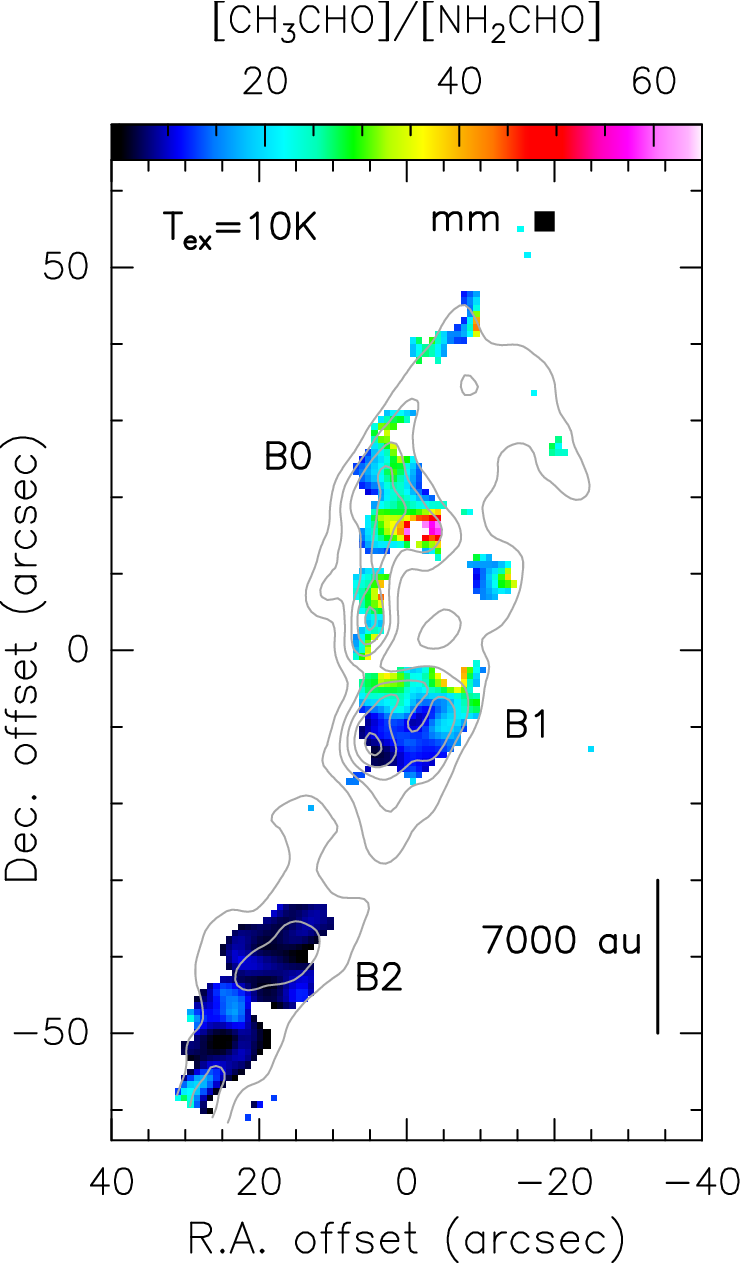}
   \caption{CH$_3$CHO/NH$_2$CHO column-density-ratio map along the L1157 southern outflow, assuming $T_\mathrm{ex}=10$\,K across the entire outflow lobe. Regions where the emission is below $3\sigma$ are masked out (the corresponding $1\sigma$ RMS values are given in the caption of Fig.\,\ref{fmaps}). The unitless colour scale is displayed at the top of the panel. Grey contours corresponding to the CS(2--1) velocity-integrated map are overlaid for reference, starting at $50\sigma$ and increasing in steps of $150\sigma$. The black square at the top of the map marks the position of L1157-mm \citep{bachiller01,lefloch17}.}
    \label{fratios10}%
    \end{figure}

For each transition listed in Table\,\ref{tlines}, we produced one column density map assuming $T_\mathrm{ex}=10$\,K across the entire outflow lobe, and another one assuming $T_\mathrm{ex}=20$\,K. We verified that the column densities computed separately for each line of a given molecule are consistent within a factor of 1.3. We then generated a CH$_3$CHO/NH$_2$CHO column density ratio map using the same molecular lines as those displayed in Fig.\,\ref{fmaps}. The ratio map for $T_\mathrm{ex}=10$\,K is shown in Fig.\,\ref{fratios10}, where we masked out the points whose line-emission intensities fall below 3$\sigma$. The column density ratio map for $T_\mathrm{ex}=20$\,K appears almost identical qualitatively and quantitatively. A quick visual inspection of these maps reveals a clear decrease in $\frac{[CH_3CHO]}{[NH_2CHO]}$ from north to south. These gradients are discussed further in Sect.\,\ref{discussion}.

The molecular column densities and column density ratios of acetaldehyde and formamide averaged over the three areas displayed in Fig.\,\ref{fmaps} (right panel) are listed in Table\,\ref{tcoldens}. The column densities we derive are of the same order as those reported in previous observational works targeting B1 with the IRAM\,30 m telescope \citep{mendoza14, lefloch17}, and with the NOEMA interferometer \citep{clau17,clau20}. The resulting values indicate that, while the gaseous column density of CH$_3$CHO remains roughly constant or marginally decreases when moving from the younger shocked region B0 to the older B2, that of NH$_2$CHO increases significantly. This confirms what could intuitively be deduced by looking at the maps in Fig.\,\ref{fmaps} and the spectra in Fig.\,\ref{fspt}. 

The aim of the following section is to explain the observed behaviours through astrochemical modelling. As the output of the astrochemical code is in terms of absolute abundances with respect to H nuclei, Table\,\ref{tcoldens} also lists these for acetaldehyde and formamide. We determined them by adopting a hydrogen column density, $N(H)$, equal to $2 \times 10^{21}$\,cm$^{-2}$ for all three shocked regions. For B1, this value is based on what is reported by \cite{lefloch12} from observations of multiple CO emission lines. For B2, we based our assumption on observations of C$^{18}$O(1--0) by \cite{Bachiller1997}, which yield similar column densities for this molecule in both B1 and B2. There are no direct measurements of CO column density reported in the literature for B0, but we adopted the same $N(H)$ value as for B1 and B2 following direct inspection of the CO(1--0) interferometric map presented by \cite{gueth96}, which shows comparable CO fluxes in both B0 and B1. It is worth mentioning that the adopted value of $N(H)$ has a high degree of uncertainty, and hence one should conservatively add a factor 2 above and below the quoted values of absolute abundances and their errors in Table\,\ref{tcoldens}. This is what we take into account in the following section.

\begin{table*}
\centering
\caption{Molecular column densities of acetaldehyde and formamide and their abundance ratios derived in B0, B1, and B2\tablefootmark{a}}
\begin{tabular}{cc|ccccc}
\hline
Position & Age\tablefootmark{b} & $N_\mathrm{CH_3CHO}$ & $N_\mathrm{NH_2CHO}$ & $X_\mathrm{CH_3CHO}$ & $X_\mathrm{NH_2CHO}$ & $\frac{[CH_3CHO]}{[NH_2CHO]}$\\
 & (yrs) & ($10^{13}$\,cm$^{-2}$) & ($10^{12}$\,cm$^{-2}$) & ($10^{-8}$) & ($10^{-10}$) & \\
 \hline
 & & \multicolumn{5}{c}{$T_\mathrm{ex} = 10$\,K}\\
\hline
B0 & 900 & $2.8\pm 0.6$ & $1.1\pm 0.2$ & $1.4\pm 0.3$ & $5.6\pm 1.2$ & $25\pm 6$\\
B1 & 1500 & $2.5\pm 0.6$ & $1.8\pm 0.3$ & $1.3\pm 0.3$ & $9.4\pm 1.4$ & $13\pm 3$\\
B2 & 2300 & $2.2\pm 0.5$ & $2.9\pm 0.4$ & $1.1\pm 0.3$ & $15\pm 2$ & $8\pm 2$\\
\hline
 & & \multicolumn{5}{c}{$T_\mathrm{ex} = 20$\,K}\\
\hline
B0 & 900 & $3.4\pm 0.7$ & $1.9\pm 0.4$ & $1.7\pm 0.4$ & $9.4\pm 2.0$ & $18\pm 5$\\
B1 & 1500 & $3.1\pm 0.7$ & $3.2\pm 0.5$ & $1.5\pm 0.3$ & $16\pm 2$ & $10\pm 2$\\
B2 & 2300 & $2.7\pm 0.7$ & $4.9\pm 0.6$ & $1.3\pm 0.3$ & $25\pm 3$ & $5\pm 1$\\
\hline
\end{tabular}
\tablefoot{
\tablefoottext{a}{The column densities are given assuming LTE and for $T_\mathrm{ex}=10$, $20$\,K (see \citealp{mendoza14,lefloch17,clau20}). Their errors are due to the flux calibration uncertainty (10\%) and the noise RMS of the associated integrated-intensity maps.}
\tablefoottext{b}{Time since the passage of the shock from \cite{podio16}, corrected for the most recent distance estimate \citep{zucker19} and rounded to the closest hundred.}
}
\label{tcoldens}
\end{table*}
    
\section{Astrochemical modelling} \label{sec:astro-model}

In this section, we describe the astrochemical modelling carried out to interpret the observations and, more specifically, the abundance ratio of acetaldehyde over formamide, [CH$_3$CHO]/[NH$_2$CHO], reported in Table\,\ref{tcoldens} and shown in Fig.\,\ref{fratios10}. We first describe the model used to simulate the passage of a shock (Sect.\,\ref{subsec:astro-model-description}), and follow with a discussion on the chemistry of CH$_3$CHO and NH$_2$CHO (Sect.\,\ref{subsec:astro-model-chemistry}).

\begin{table}
    \centering
        \caption{Parameters used to model the passage of the shock (Sect. \ref{sec:astro-model}).}
    \begin{tabular}{l|c|c}
        \hline
        \hline
        \multicolumn{3}{c}{Physical parameters of the shocked gas} \\
        Parameter & L1157-B1 value & Range\\
        \hline
        n$_{H_2}$    [cm$^{-3}$] & $4\times 10^5$     & 2--8 $\times 10^5$ \\
        T            [K]         & 90                 &  50--150 \\
        $t_{shock}$ [yr]         &  1600              & 200-3000 \\
        $\zeta_{CR}$ [s$^{-1}$]  & $6\times 10^{-16}$ &  3--12 $\times 10^{-16}$\\
        \hline
        \hline
        \multicolumn{3}{c}{Abundances (wrt H) of the injected species} \\
        Species & L1157-B1 value & Range\\
        \hline
        H$_2$O    & $2\times 10^{-4}$ & 1--4 $\times 10^{-4}$\\
        CO$_2$    & $3\times 10^{-5}$ & 1.5--6 $\times 10^{-5}$\\
        CH$_3$OH  & $4\times 10^{-6}$ & 2--8 $\times 10^{-6}$\\
        NH$_3$    & $2\times 10^{-5}$ & 1--4 $\times 10^{-5}$\\
        H$_2$CO   & $1\times 10^{-6}$ & 0.5--2 $\times 10^{-6}$\\
        OCS       & $2\times 10^{-6}$ & 1--4 $\times 10^{-6}$\\
        SiO       & $1\times 10^{-6}$ & 0.5--2 $\times 10^{-6}$\\
        Si        & $1\times 10^{-6}$ & 0.5--2 $\times 10^{-6}$\\
        CH$_3$CH$_2$ & $4\times 10^{-8}$   & 0--20 $\times 10^{-8}$\\
        CH$_3$CH$_2$OH & $6\times 10^{-8}$ & 0--40 $\times 10^{-8}$\\
        \hline
        CH$_3$CHO & 0 & 0--4 $\times 10^{-8}$\\
        NH$_2$CHO & 0 & 0--4 $\times 10^{-9}$\\
        \hline
    \end{tabular}
    \tablefoot{
    The upper half table lists the adopted H$_2$ density, n$_{H_2}$, temperature, $T$, time since the shock passage, $t_{shock}$, and the cosmic-ray ionisation rate, $\zeta_{CR}$, of the gas after the passage of the shock. The lower half of the table lists the adopted abundances of the species injected into the gas phase from the grain mantle after the passage of the shock. The values quoted in the central column are the results of previous modelling of the L1157-B1 shocked gas recently discussed in \cite{Tinacci2023-gretobapegas} and \cite{Giani2023-ch3cn}. Column 3 lists the range of each parameter used in the modelling to take into account the differences in B0 and B2.
    }
    \label{tab:model-parameters}
\end{table}

\subsection{Model description} \label{subsec:astro-model-description}

We used the code \textsc{GRAINOBLE} \citep{Taquet2012-grainoble,Ceccarelli2018-grainoble} with the gas-phase-only option, which allows us to follow the chemical evolution of the gas with time.
In order to describe the impact on the chemistry after the passage of a shock, we followed a scheme described in previous works by our group \citep[e.g.][]{podio14,clau17,Tinacci2023-gretobapegas, Giani2023-ch3cn}.
Briefly, the passage is modelled as a sudden increase in the gas density and temperature, as well as in the gas abundances of species that are known to be mostly frozen onto the grain mantles and liberated into the gas phase at the passage of the shock. Therefore, the model consists of two steps:

\begin{itemize}
\item \textit{Step 1,} where we compute the abundances of a typical molecular cloud ($T_\mathrm{gas}\,=$\,10 K and $n_{H_2}=2\times 10^4$ cm$^{-3}$) at steady state. This step runs for a sufficiently long time ($10^8$\,yr) to reach stable abundances typical of cold molecular clouds.
\item \textit{Step 2}, where the gas molecular abundances at the end of Step 1 are used as inputs. In this second step, we follow the chemical evolution of the gas with higher values of temperature, density, and ice-component gas-phase abundances, as indicated in Table\,\ref{tab:model-parameters}. This is done to artificially simulate the passage of a shock, which causes a sudden gas temperature and density increase, as well as an immediate release of icy molecules into the gas. The model then lets the gas chemistry evolve in these physical conditions, that is, without varying temperature, density, or cosmic-ray ionisation rate, for about 3000\,yr.
\end{itemize}

Table\,\ref{tab:model-parameters} summarises the characteristics of the shocked gas towards L1157-B1, which has been extensively studied and modelled by our group \citep[for the description of the most recently adopted values, see][]{Tinacci2023-gretobapegas, Giani2023-ch3cn}. In order to model the other shocked regions of L1157, B0, and B2, we varied the H$_2$ density, n$_{H_2}$, and cosmic-ray ionisation rate, $\zeta_{CR}$, by a factor of two above and below the B1 values to estimate their impact on the chemistry. Similarly, we ran models with gas temperatures ranging between 50 and 150 K. We also considered models with injected species abundances that vary by a factor of two above and below those of B1. CH$_3$CH$_2$ and CH$_3$CH$_2$OH are exceptions to this, in the sense that a wider range of abundances was considered in order to better explore the different gas-phase reactions leading to CH$_3$CHO as explained in Sect.\,\ref{subsec:astro-model-chemistry}. As discussed in Sect.\,\ref{disc2} and listed in the last two lines of the table, we also explored a range of injected abundances of formamide and acetaldehyde from the grain surfaces.

\subsection{Chemistry of CH$_3$CHO and NH$_2$CHO} \label{subsec:astro-model-chemistry}

In this work, we consider two major observables: acetaldehyde and formamide. In the following, we briefly recall the chemistry of each of these species and what is known about the crucial molecules involved in it in the L1157 southern outflow. However, as our observations coupled with our modelling are only able to constrain the gas-phase reaction pathways, Table \ref{tab:model-reaction-list} only summarises the gas-phase reactions involved in the formation of acetaldehyde and formamide. 

\begin{table}
    \centering
    \caption{Gas-phase reactions forming acetaldehyde and formamide.}
    \begin{tabular}{clll}
    \hline
      No. & Reactants   & Products & Ref. \\
      \hline
      \hline
      \multicolumn{4}{c}{Acetaldehyde (CH$_3$CHO)}\\
       1  & CH$_3$CH$_2$ + O & CH$_3$CHO + H & 1 \\
       2a & CH$_3$CH$_2$OH + OH & CH$_3$CHOH + H$_2$O & 2\\
       2b & CH$_3$CHOH + O &  CH$_3$CHO + OH & 2\\
       \hline
      \multicolumn{4}{c}{Formamide (NH$_2$CHO)}\\
      3  & H$_2$CO + NH$_2$ & NH$_2$CHO + H & 3 \\
      \hline
    \end{tabular}
    \tablebib{(1)~\citet{Charnley2004A-ch3cho}; (2) \citet{skouteris18}, (3) \citet{skouteris17}.}
    \label{tab:model-reaction-list}
\end{table}
%

\subsubsection{Acetaldehyde}

The formation route of acetaldehyde is debated.
Two major routes have been proposed in the literature: in the gas phase \citep[e.g.][]{Charnley2004A-ch3cho, vastel14, vazart20} and on the grain surfaces \cite[e.g.][]{garrod08,joan21,garrod22,Ibrahim2022-acetaldehyde}. In our modelling, the amount of acetaldehyde injected from the grain mantles is treated as a parameter that is independent of the possible mechanism operating on the grain surfaces. This is because our observations are not able to put constraints on the specific surface-chemistry routes, as what matters is solely the amount of injected acetaldehyde in the gas phase.
On the other hand, our observations are potentially able to constrain the gas-phase routes of acetaldehyde formation, if any, because its abundance depends on the abundances of the  reactants and the age of the shocked region.
This last parameter is particularly important because gas-phase reactions take time to produce acetaldehyde \citep[see e.g.][]{clau17}.
Therefore, in what follows we only review the gas-phase formation reactions of acetaldehyde.

Several gas-phase reaction routes have been invoked in the literature. 
\cite{vazart20} carried out a systematic and careful review of all of them, and performed new quantum-mechanics (QM) computations when the reaction rate constants and/or branching ratios had not been previously characterised, especially at the cold (10--200 K) temperatures typical of the molecular ISM.
These authors found that only two reactions are feasible at such temperatures:
\textit{(1)} the reaction of ethyl radical (CH$_3$CH$_2$) with oxygen, and \textit{(2)} a  two-reaction chain starting from ethanol (CH$_3$CH$_2$OH), as summarised in Table\,\ref{tab:model-reaction-list}.

The acetaldehyde abundance predicted by these two routes depends on the abundance of ethyl radical and oxygen, and that of ethanol, O and OH, respectively.
The atomic oxygen and OH abundances increase after the passage of the shock, because of the injection of icy water into the gas-phase followed by further reactions. Therefore, the O and OH abundances are self-consistently computed by the code.
On the other hand, ethyl radical and ethanol are injected directly from the grain mantles, because they are possibly grain-surface products \citep{Perrero2022-ethanol, McClure2023-ices}.
As direct measurements of their abundances are not available in the literature, except for a work in progress for ethanol (Robuschi et al. \textit{in prep.}), they are considered as free parameters and we varied them from 0 to 20 and 4 $\times 10^{-7}$, as indicated in Table\,\ref{tab:model-parameters}.

\subsubsection{Formamide}

As in the case of acetaldehyde, the formation route of formamide has been a source of debate since its first detection towards a solar-like protostar \citep{kahane13}.
Several possibilities have been proposed and discussed in the literature, which are reviewed in \cite{ls19}. 
We therefore limit the present discussion to studies published after this review, following a brief summary of previous work. 
As mentioned above for acetaldehyde, our observations can only constrain the gas-phase routes leading to the formation of formamide. 
We thus briefly comment on the grain-surface routes, for completeness, but dedicate the greater part of our discussion to the gas-phase routes.

\paragraph{Grain-surface formation}

According to the literature, there are primarily three reaction pathways leading to formamide on the surface of dust grains.

\textit{(1)} Hydrogenation of frozen isocyanic acid (HNCO) was favoured in several studies based on the tight correlation found between the abundance of this molecule and that of formamide (e.g. \citealp{mendoza14,ls15}). 
However, the process of hydrogenation of solid HNCO was found not to be viable in laboratory-based experiments \citep{noble15, Haupa2022}, which is why alternative pathways were explored after the publication of these results (see below). 

\textit{(2)} The combination of frozen amino  (NH$_2$) and formyl (HCO) radicals on the grain surfaces, as proposed in the modelling work by \cite{garrod08}, was studied by \cite{joan19} and \cite{Enrique-Romero2022} using quantum chemistry computations on a cluster of amorphous solid water molecules. 
The authors found that formamide is one of the possible products of the reaction, but that it competes with the channel leading to CO + NH$_3$ (see also \citealp{rimola18}).

\textit{(3)} The reaction of a frozen water molecule of the ice mantle with a landing cyano radical (CN) was studied by \citet{rimola18}. These authors found that solid water molecules act as catalytic active sites that facilitate the H transfers involved in the process, thus reducing the energy barriers with respect to the analogous reaction in the gas phase.

\paragraph{Gas-phase formation}
The gas-phase neutral--neutral reaction H$_2$CO + NH$_2$ has been thoroughly studied theoretically in the past few years. 
Until recently, QM computations of this reaction existed, with some discordant results that merit discussion. \cite{barone15} first proposed that the reaction has an embedded energy barrier that leads to a rate constant at 100 K equal to $4.3\times 10^{-11}$ cm$^3$s$^{-1}$.
The embedded energy barrier was challenged by \cite{song16} and revised by \cite{vazart16} and \cite{skouteris17}, who predicted a rate constant at 100 K equal to $1.2\times 10^{-13}$ cm$^3$s$^{-1}$.

Very recently, \cite{douglas22} experimentally studied this reaction at 34 K using reaction kinetics in uniform supersonic flow and were not able to detect formamide, setting an upper limit to the rate constant equal to $6\times 10^{-12}$ cm$^3$s$^{-1}$, which is 3.3 times larger than that predicted by \cite{skouteris17} at the same temperature ($1.8\times 10^{-12}$ cm$^3$s$^{-1}$).
\cite{douglas22} complemented the experimental results with new QM computations, the accuracy of which is essentially the same as for those mentioned above \citep{song16, vazart16, skouteris17}.
However, \cite{douglas22} insist on the presence of a non-embedded activation barrier, which would lead to a much lower rate constant at temperatures of $\leq 200$ K.
The presence or absence of the barrier is linked to how the pre-reactant complex (PRC) is computationally treated and to whether or not the zero-point energy (ZPE) should be added to the transition state (TS) towards the formation of formamide.
The first group of authors \citep{barone15, vazart16,skouteris17} claim that the ZPE, computed with the standard methods adopted by all the cited authors, should not be added to the PRC TS energy because the PRC has very loose modes, which would not substantially alter the TS energy height.
The second group \citep{song16,douglas22} believes the opposite to be true, and these authors consider the ZPE of the PRC computed in the harmonic approximation.
As the various approximations used for the computations of the standard TS very likely do not apply to the PRC, at this stage, it is impossible to confirm with certainty which of these positions is correct and whether or not the ZPE of the PRC could lead to an activation barrier for the H$_2$CO + NH$_2$.
New calculations with more adapted methods should be employed or a lower upper limit in experimental works should be obtained.
In summary, the new experimental work by \cite{douglas22} does not bring a significant constraint when compared to the computations carried out by \cite{vazart16} and \cite{skouteris17}.

Given this picture and the fact that the value reported by \cite{skouteris17} provided model predictions in agreement with the observations by \cite{clau17}, in the present work, we adopt the \cite{skouteris17} rate constants which, we emphasise, are compatible with the experimental results of \cite{douglas22}.


\section{Discussion}\label{discussion}

\subsection{Results of the gas-phase astrochemical modelling} \label{subsec:astro-model-results}

\begin{figure}[!htb]
   \centering
   \includegraphics[scale=0.53]{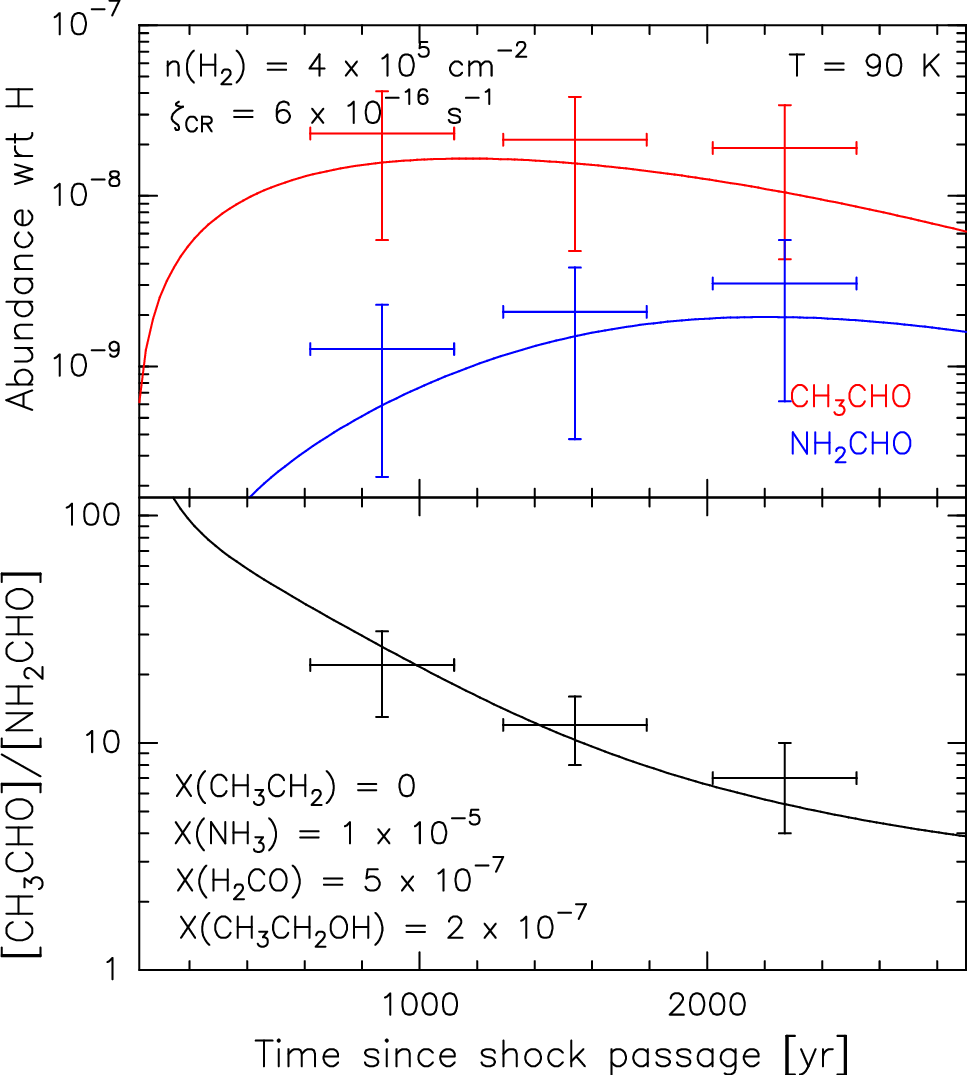}
   \caption{Solid lines show the CH$_3$CHO and NH$_2$CHO abundances (\textit{top panel}) and their ratio (\textit{bottom panel}) as predicted by the gas-phase chemistry model that best reproduces our observations in L1157-B. 
   The physical parameter values, as well as the abundances of key reactants injected into the gas phase after the shock passage, are indicated in the upper and bottom panels, respectively. 
   Measurement points in B0, B1, and B2 are shown as crosses whose vertical lengths correspond to the measured abundance and ratio uncertainties (see Table\,\ref{tcoldens}), and whose horizontal lengths account for a 500 yr uncertainty in the post-shock time as derived by \cite{podio16}.}
              \label{fmodgas}%
    \end{figure}

As a first step, we explored the model predictions, varying the injected abundances (bottom half of Table\,\ref{tab:model-parameters}) while keeping the physical parameters 
fixed (upper half of Table\,\ref{tab:model-parameters}). At this stage of the discussion, we consider no injection of either formamide or acetaldehyde from the dust grains. We ran 42 models in total. This allowed us to evaluate how the abundances of formamide and acetaldehyde vary with time when changing the initial injected abundances of the reactants leading to their formation in the gas (Table\,\ref{tab:model-reaction-list}) as well as other potentially relevant species. We then identified the model that best reproduces the observations presented in the previous sections, both in terms of absolute abundances and their ratio. Figure\,\ref{fmodgas} shows the predictions of this `best-fit' model as a function of time along with the observed values. We emphasise here that  [CH$_3$CHO]/[NH$_2$CHO] places more stringent constraints on the modelling than their respective absolute values, given the high uncertainty associated with $N(H)$, as explained in Sect.\,\ref{ratio}.

In practice, the `best-fit' model has all the reference values of Table\,\ref{tab:model-parameters} (i.e. the values listed in the central column), with the following exceptions:

\begin{enumerate}
\item The injected abundances of ammonia (NH$_3$) and formaldehyde (H$_2$CO) are $1\times10^{-5}$ and $5\times 10^{-7}$, which are lower by a factor of 2 than the reference values listed in Table\,\ref{tab:model-parameters} for B1. Larger abundances overestimate formamide production.  Interestingly, this requirement better fits the recent direct measurements of $X(NH_3)$ reported by \cite{feng22} along the southern outflow, which are on the order of $(1-4)\times 10^{-6}$. This abundance range is consistent within a factor of 2 with the modelled values at the age of B1 and B2, which lie in the range $(5-7)\times10^{-6}$. The best-fit model also agrees with the abundances of H$_2$CO derived in B1 by \cite{fontani14}, and in B1 and B2 by \cite{Bachiller1997}, which are lower than $10^{-6}$.
\item The abundance of ethanol (CH$_3$CH$_2$OH) is in the range $(2-4)\times 10^{-7}$, which is more than a factor of 2 larger than the reference value. This is consistent with the measured values of $\sim10^{-8}$ for this species (\citealp{lefloch17}, Robuschi et al. \textit{in prep.}), as the model output shows that this molecule is consumed relatively quickly over time, reaching the observed abundance after about 1500\,yr. Regarding ethyl radical (CH$_3$CH$_2$), any abundance between 0 and $1\times 10^{-7}$ appears to provide a reasonable fit to the data. In Fig.\,\ref{fmodgas} we present the results from the model with no injected CH$_3$CH$_2$.
\end{enumerate}

As is always the case with numerical models, it is important to be aware of their caveats and limitations.  In the case of the present study, the main caveat of our astrochemical modelling is that the physical properties do not vary with time, or along the outflow, which is likely unrealistic.  This is why, as a second step in our analysis, we dedicated further effort to exploring the physical parameter space of our modelling. To this aim, we took the injected abundances of the `best-fit' model as a new reference, and varied the density, gas temperature, and CR ionisation rate around the reference values in order to explore how the chemistry related to acetaldehyde and formamide varies with these physical parameters. 
The results from this analysis are illustrated in Fig.\,\ref{fmodphys}. In summary, they show that, while the abundance of acetaldehyde is barely affected by changes in gas density and/or temperature, formamide abundance increases with increasing density and with decreasing gas temperature. The effect of CR ionisation rate is similar on both molecular species, with higher values essentially accelerating molecular destruction with respect to lower values. This analysis allows us to visualise the chemical outcome resulting from different physical parameters for the different shocked regions. For example, if the gas in B2 is colder than that in B0 and B1, which might be plausible given the longer time elapsed since it was shocked, then we would need to decrease the amount of injected formaldehyde and ammonia for this region in order to better reproduce its formamide abundance.

\subsection{Formamide: Gas or grain-surface chemistry product?}\label{disc2}

\begin{figure}[!htb]
   \centering
   \includegraphics[scale=0.53]{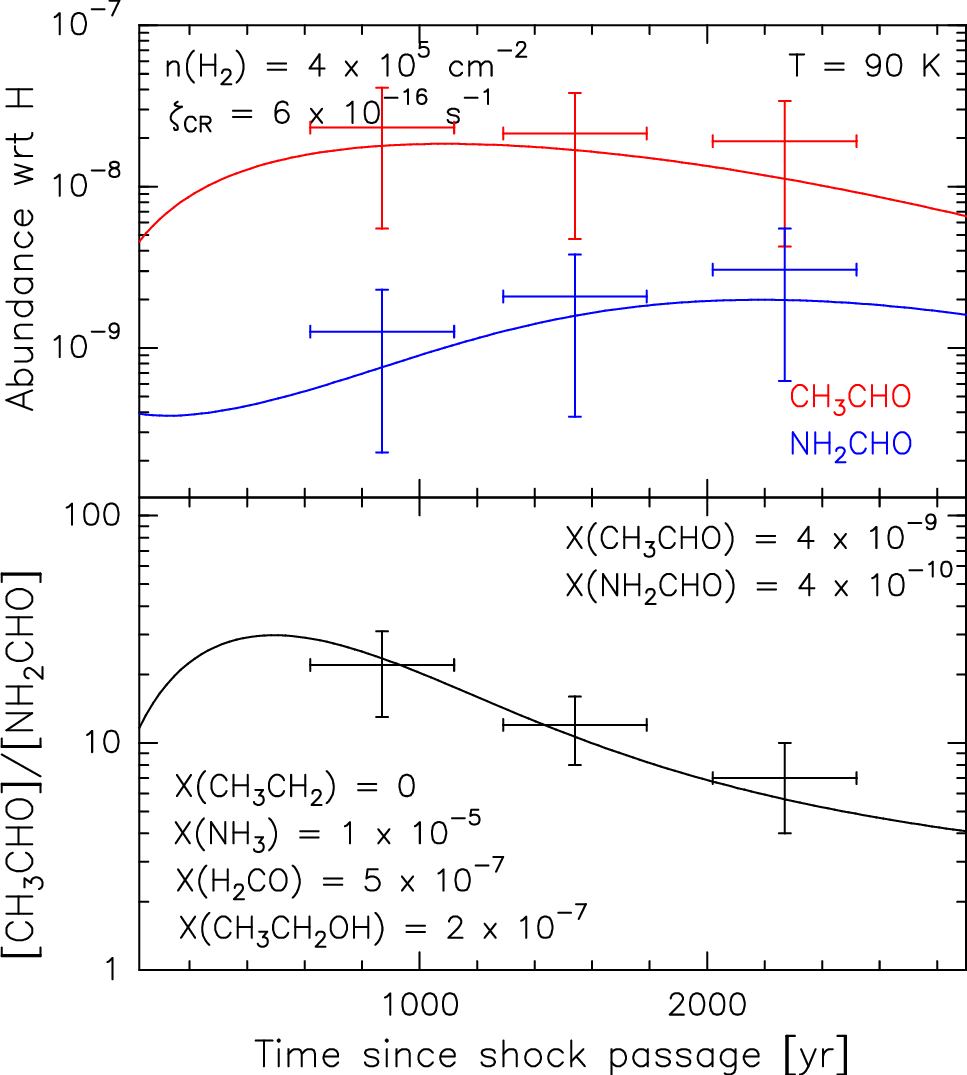}
   \caption{CH$_3$CHO and NH$_2$CHO abundances (\textit{top panel}) and their ratio (\textit{bottom panel}) predicted by the gas-phase chemistry model that best represents our observations in L1157-B, including injection of formamide and acetaldehyde from the grains into the gas phase at the shock passage as labelled. As in Fig.\,\ref{fmodgas}, we indicate the physical parameter values, as well as the abundances of these molecules' key reactants injected into the gas. Measurement points in B0, B1, and B2 are shown as crosses whose vertical lengths correspond to the measured abundance (ratio) uncertainties (see Table\,\ref{tcoldens}), and whose horizontal lengths account for a 500 yr uncertainty in the post-shock time as derived by \cite{podio16}.}
              \label{fmodgrain}%
    \end{figure}

The acetaldehyde-over-formamide abundance ratio we have measured along the southern L1157 outflow lobe follows a net decrease from B0 to B2. The results from our gas-phase astrochemical modelling show that there is no need for grain-surface chemistry leading to either formamide or acetaldehyde in order to reproduce our observations. At the same time, these results do not necessarily exclude a contribution from grain-surface chemistry. For this reason, we ran a total of 16 extra models in which varying amounts of these two species are released from the grains into the gas after the shock's passage (see the two bottom lines in Table\,\ref{tab:model-parameters}). As an example, Fig.\,\ref{fmodgrain} shows the result of injecting formamide and acetaldehyde abundances of $4\times 10^{-10}$ and $4\times 10^{-9}$, respectively, from the dust grains, followed by gas-phase chemistry. As can be seen, this model also reproduces  our observations well. In other words, surface chemistry cannot be excluded as an extra means to produce formamide and acetaldehyde. Indeed, we find that the observational values are well reproduced within the uncertainties as long as the respective injected abundances of acetaldehyde and formamide do not exceed $4\times 10^{-8}$ and $4\times 10^{-9}$ relative to H nuclei. These two values are strict upper limits, meaning that any higher injected abundance will systematically overestimate the observed absolute abundances in the gas. However, they are not independent of each other in terms of the resulting abundance ratio. Indeed, for the maximum possible value of injected acetaldehyde abundance, $4\times 10^{-8}$, the injected amount of formamide must be in the range $(1-4)\times 10^{-9}$. In other words, the ratio of acetaldehyde/formamide injected abundances must be between 10 and 40 to reproduce the observed values and any ratio below 10 fails to reproduce them.

In conclusion, surface chemistry cannot be excluded as an extra means to produce formamide and acetaldehyde. However, completely switching off the gas-phase reactions listed in Table\,\ref{tab:model-reaction-list} does not allow us to reproduce our observations. Indeed, in a grain-surface-only chemistry scenario, one would expect the abundance of gaseous formamide to either remain constant or decrease with time due to destruction processes, which is the opposite of what we observe. Therefore, gas-phase chemistry is a necessary ingredient to match what we observe in this region. This agrees with what was already concluded by \cite{clau17}, where the authors show that switching off the gas-phase reactions leading to formamide and acetaldehyde yields an increasing [CH$_3$CHO]/[NH$_2$CHO] ratio with time, contrasting with our decreasing trend.

As a final note, it is worth adding that the measured [CH$_3$CHO]/[NH$_2$CHO] ratio assumes co-spatial emission of acetaldehyde and formamide in the line-of-sight direction, which is not necessarily true. This may be the case for B1, in  particular, as it is known to be highly complex in terms of morphology and kinematic signatures (e.g. \citealp{benedettini07,lefloch12}), which is altogether suggestive of several jet impacts having occurred on this cavity at different times. This might explain the spread in values both for the individual NH$_2$CHO and CH$_3$CHO column densities and the [CH$_3$CHO]/[NH$_2$CHO] ratio in B1, as seen in Fig.\,\ref{fratios10}. In this respect, the inclusion of the B0 and B2 cavities in our study has proven to be a valuable addition with respect to the B1-only study performed
by \cite{clau17}, allowing us to better understand what chemical pathways dominate in the formation of formamide in shocked regions. We therefore conclude that (i) gas-phase chemistry works to produce interstellar formamide and is necessary to reproduce our observations; (ii) a contribution from grain-surface chemistry is also possible and cannot be excluded; and (iii) surface chemistry alone cannot account for the observed amount of formamide in the L1157 southern outflow lobe. On the other hand, our modelling results do not allow us to distinguish between gas and grain-surface chemistry regarding the formation of acetaldehyde. This was already mentioned by \cite{clau20}, who rightly suggested that, given the rapid gas-phase chemistry leading to acetaldehyde, much younger shocked regions should be explored in order to better discriminate between gas and grain-surface chemistry, which provides an avenue for future studies.
\section{Conclusions}\label{conclusions}

We used the IRAM NOEMA interferometer to map the NH$_2$CHO and CH$_3$CHO line emission at 3\,mm and 4-5$"$ resolution along the entire southern outflow lobe driven by the L1157 protobinary. Our map therefore covers the three shocked regions B0, B1, and B2, each successively older than the previous one. Our main results are summarised as follows.

   \begin{enumerate}
      \item We detected, above a $S/N=5$, three NH$_2$CHO and two CH$_3$CHO lines, all with similar upper-level energies. While the former emits most strongly in the older B2 cavity, the latter shows a similar intensity in all three cavities.
      \item Assuming LTE and a homogeneous excitation temperature along the outflow ($T_\mathrm{ex}=10$\,K), we derive a [CH$_3$CHO]/[NH$_2$CHO] ratio ranging between 25 in B0 and 8 in B2. The abundance ratio between the two molecules therefore decreases significantly between the youngest and the oldest shocked region.
      \item The results from our astrochemical modelling favour two possible scenarios: one in which the amount of formamide observed can be explained by gas-phase-only chemistry, more specifically via the reaction H$_2$CO + NH$_2$ $\rightarrow$ NH$_2$CHO + H$_2$, and another scenario in which part of the observed formamide originates from surface chemistry and part from gas-phase chemistry. Solid-chemistry alone cannot account for the [CH$_3$CHO]/[NH$_2$CHO] ratio measured along the L1157 outflow lobe.
      \item Our observations do not allow us to distinguish whether it is gas or grain-surface chemistry that dominates in the formation of acetaldehyde.
    \end{enumerate}

The present study highlights the importance of exploring shocked regions with different post-shock ages in order to gain better insight into the evolution of gas-phase organic chemistry in such regions. Our findings indicate that gas chemistry plays an important role in the synthesis of the prebiotic precursor formamide.

\begin{acknowledgements}
We would like to warmly thank our anonymous referee for providing helpful suggestions that added value and clarity to this manuscript. We thank Nadia Balucani, Layal Chahine, Lisa Giani, and Fr\'ed\'erique Motte for their participation in fruitful discussions during the preparation of this paper. We are grateful to the entire IRAM-NOEMA staff for their precious support during the various observation runs. This project has received funding from: 1) the European Research Council (ERC) under the European Union's Horizon 2020 research and innovation program, for the Project “The Dawn of Organic Chemistry” (DOC), grant agreement No 741002; 2) the PRIN-INAF 2016 The Cradle of Life - GENESIS-SKA (General Conditions in Early Planetary Systems for the rise of life with SKA); 3) the European Union’s Horizon 2020 research and innovation programs under projects “Astro-Chemistry Origins” (ACO), Grant No 811312. L. Podio and C. Codella acknowledge financial support under the National Recovery and Resilience Plan (NRRP), Mission 4, Component 2, Investment 1.1, Call for tender No. 104 published on 2.2.2022 by the Italian Ministry of University and Research (MUR), funded by the European Union – NextGenerationEU-Project Title 2022JC2Y93 Chemical Origins: linking the fossil composition of the Solar System with the chemistry of protoplanetary disks – CUP J53D23001600006 – Grant Assignment Decree No. 962 adopted on 30.06.2023 by the Italian Ministry of Ministry of University and Research (MUR). L. Podio and C. Codella also acknowledge the PRIN-MUR 2020 BEYOND-2p (Astrochemistry beyond the second period elements, Prot. 2020AFB3FX), the project ASI-Astrobiologia 2023 MIGLIORA (Modeling Chemical Complexity, F83C23000800005), the INAF-GO 2023 fundings PROTO-SKA (Exploiting ALMA data to study planet forming disks: preparing the advent of SKA, C13C23000770005), and the INAF Mini-Grant 2022 “Chemical Origins” (PI: L. Podio).
\\
\end{acknowledgements}

\bibliographystyle{aa}
\bibliography{references.bib}

\begin{appendix}

\section{Comparison of IRAM 30 m and NOEMA spectra}\label{appa}

\begin{figure}[!htb]
   \centering
   \includegraphics[scale=0.5]{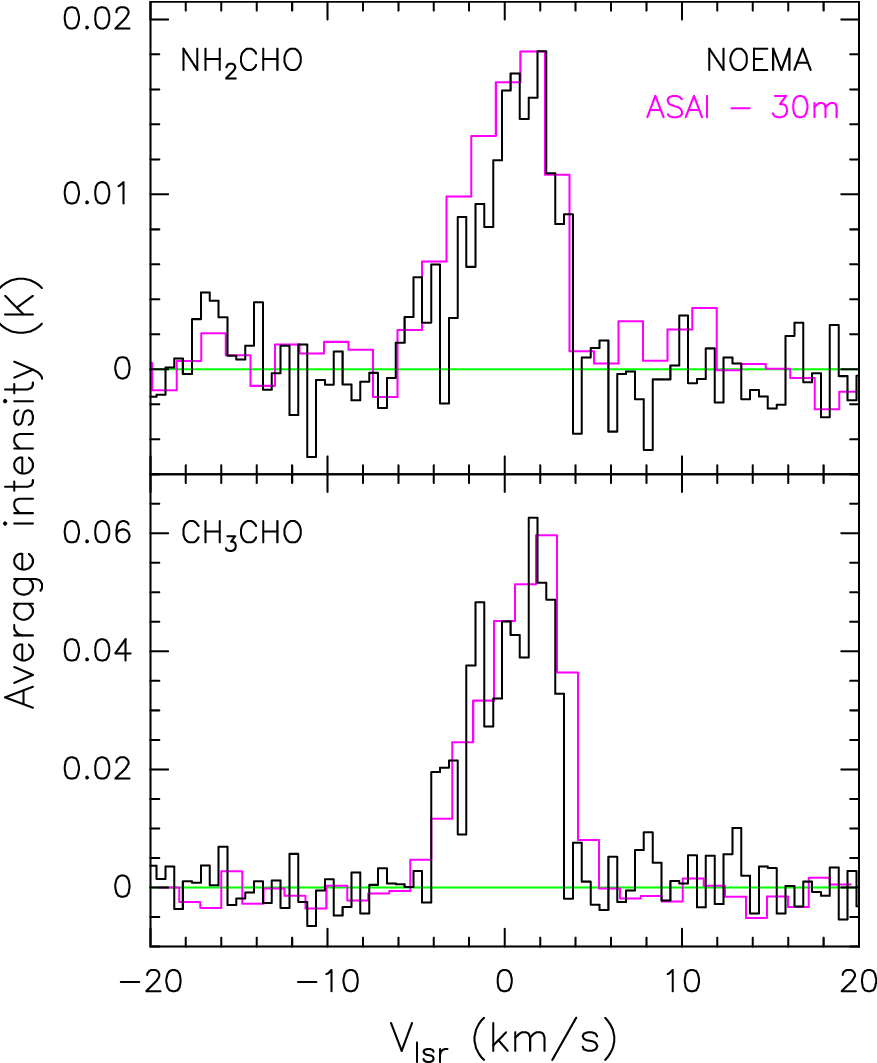}
   \caption{IRAM\,30 m spectra of CH$_3$CHO(5$_{1,4}-4_{1,3}$)A and NH$_2$CHO(4$_{0,4}-3_{0,3}$) in L1157-B1 obtained within the context of ASAI (magenta), with the respective NOEMA spectra averaged over the IRAM\,30 m beam (this work; black).}
              \label{floss}%
    \end{figure}

Figure\,\ref{floss} shows a comparison between the formamide and acetaldehyde spectra obtained with the IRAM\,30 m in the context of the ASAI Large Program \citep{lefloch18} and our NOEMA spectra, averaged over an area equal to the 30 m half-power beam size. The spectra displayed correspond to the L1157-B1 shocked region. It is clear that most of the emission is recovered by the interferometer for these two lines.

\section{Constraints on $T_\mathrm{rot}$ and dependence of molecular column densities on $T_\mathrm{ex}$}\label{appb}

\begin{figure}[!htb]
   \centering
   \includegraphics[scale=0.6]{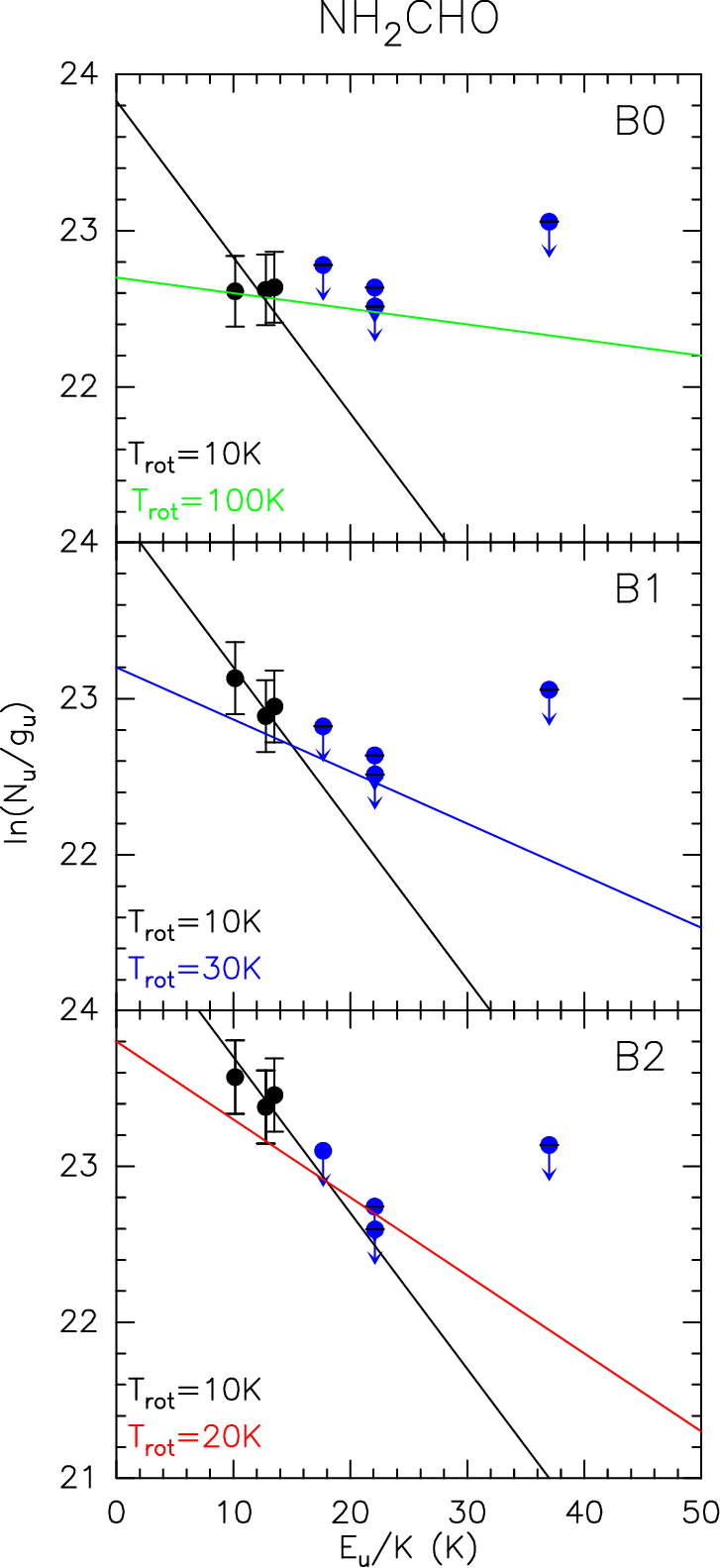}
   \caption{Rotational diagrams of NH$_2$CHO for B0, B1, and B2, including 3$\sigma$ upper limits of undetected transitions in blue. Several curves corresponding to different values of $T_\mathrm{rot}$ are plotted for reference only and do not correspond to actual fits to the data.}
              \label{frd}%
    \end{figure}

The molecular transitions detected for both formamide and acetaldehyde have similar upper-level energies ($E_\mathrm{up}$), preventing us from accurately constraining the rotational temperature, $T_\mathrm{rot}$, associated with each shocked region. However, we have used the 3$\sigma$ upper limits of four undetected formamide transitions spanning $E_\mathrm{up}$ values between 17 and 37\,K to try to constrain $T_\mathrm{rot}$ through a rotational diagram plot. The results are shown in Fig.\,\ref{frd}, together with several curves corresponding to different values of $T_\mathrm{rot}$. The latter are not actual fits to the data but merely a tool to help us constrain the rotational temperature, which is the inverse of the curve's slope. We conclude that, while $T_\mathrm{rot}$ cannot be accurately constrained for B0, since it can take any value above 10\,K, B1 and B2 need to have values below 30 and 20\,K, respectively. Unfortunately, no useful undetected transitions of CH$_3$CHO were found in our dataset to constrain $T_\mathrm{rot}$ for this species, since their associated spontaneous de-excitation coefficients, $A_{ij}$, are too low to place any meaningful constraints.

This analysis, coupled with the findings of previous single-dish results, allows us to be confident in our choice of $T_\mathrm{ex} = 10-20$\,K, at least for B1 and B2. Indeed, \cite{mendoza14} used multiple lines of NH$_2$CHO observed with the IRAM\,30 m data on both B1 and B2 to derive $T_\mathrm{rot} = 10$\,K in these two shocked regions. As for CH$_3$CHO, \cite{clau17} derived $T_\mathrm{rot} = 8$\,K for B1 from NOEMA observations at 2\,mm wavelength, while \cite{lefloch17} estimated $T_\mathrm{rot} = 17$\,K for B1 from IRAM\,30 m data. Our adopted range of $T_\mathrm{ex}$ is therefore reasonable. Nevertheless, it is still subject to uncertainty, which is why in this section we aim to evaluate how the molecular column densities and [CH$_3$CHO]/[NH$_2$CHO] change for different values of $T_\mathrm{ex}$.

   \begin{figure}[!htb]
   \centering
   \includegraphics[scale=0.5]{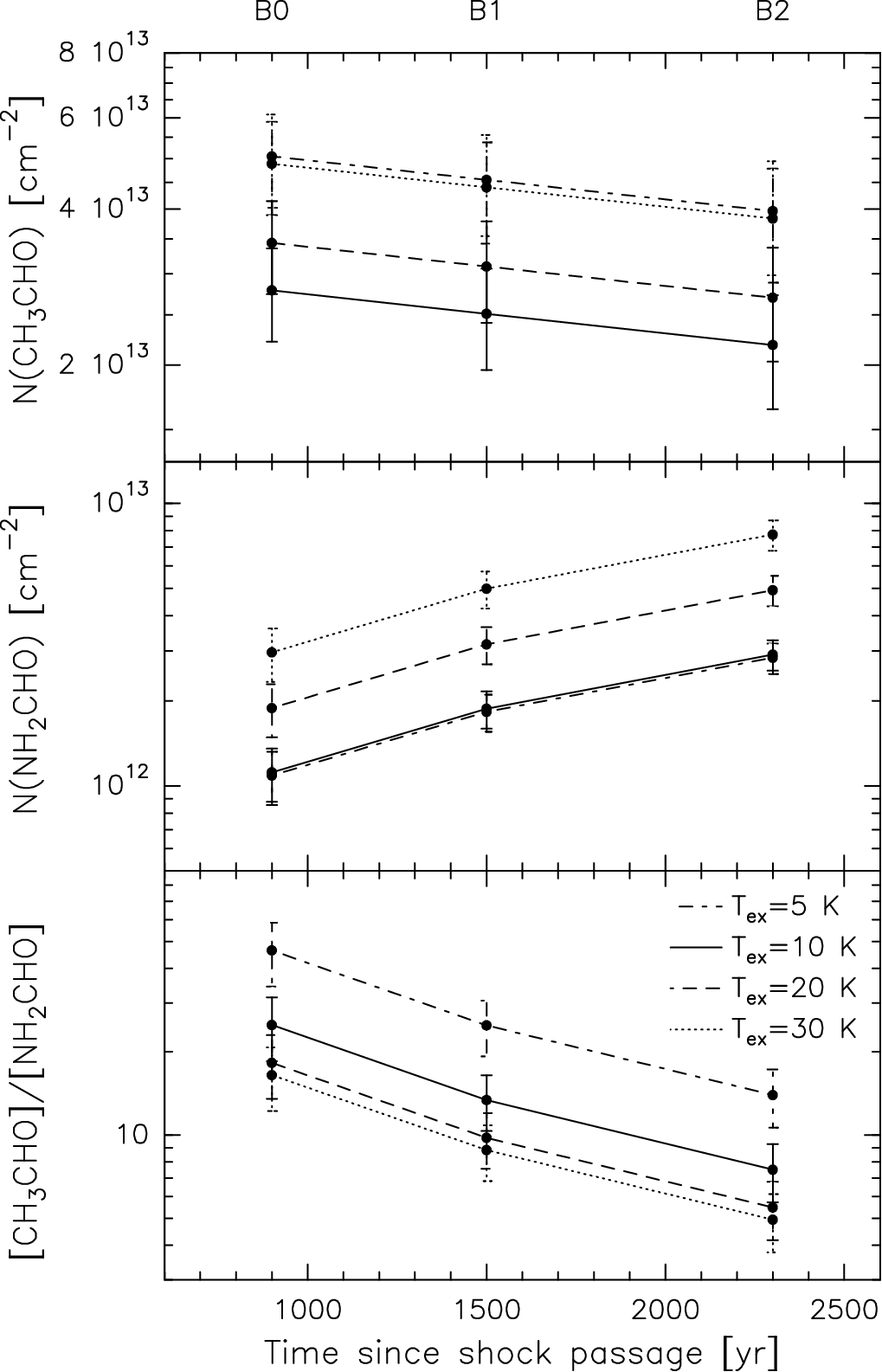}
   \caption{Molecular column densities of CH$_3$CHO and NH$_2$CHO (\textit{top panel}) and their abundance ratio (\textit{bottom panel}) against time since the passage of the shock, for four different values of excitation temperature, $T_\mathrm{ex}$. The times associated with each shocked region are based on \cite{podio16}.}
              \label{fratiosall}%
    \end{figure}

Figure\,\ref{fratiosall} illustrates the variation of column density of acetaldehyde and formamide, as well as their ratio, for four different values of $T_\mathrm{ex}$, ranging between 5 and 30\,K. It can be seen that the CH$_3$CHO column density varies at most by a factor 2 within this temperature range, while the NH$_2$CHO column density changes by less than a factor 3. On the other hand, the ratio of these two column densities, i.e. what our astrochemical modelling aims to reproduce, does not change significantly within the range $T_\mathrm{ex} = 10-20$\,K, which is the temperature interval that best matches previous observational studies.

\section{Dependence of the astrochemical modelling on physical parameters}\label{appc}

   \begin{figure*}[!htb]
   \centering
   \begin{tabular}{lcr}
   \includegraphics[scale=0.42]{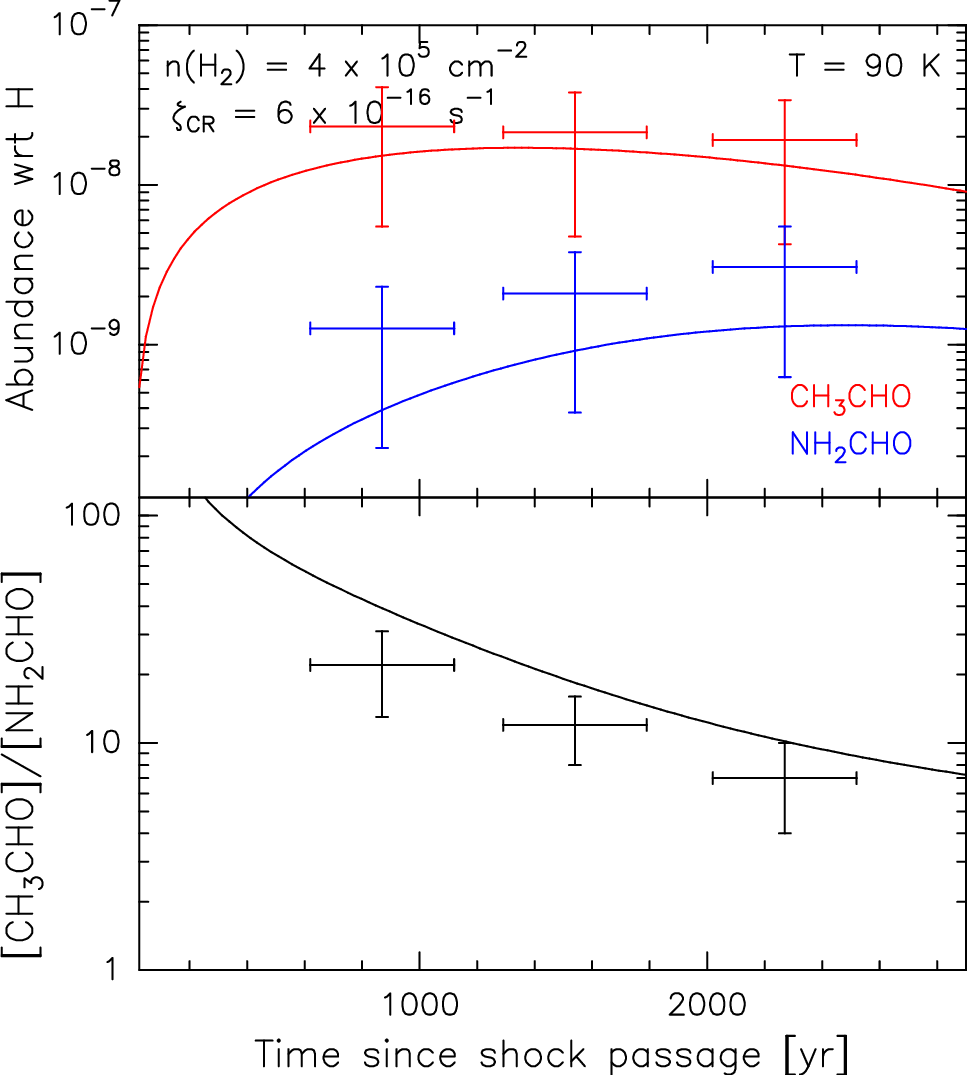} & & \includegraphics[scale=0.42]{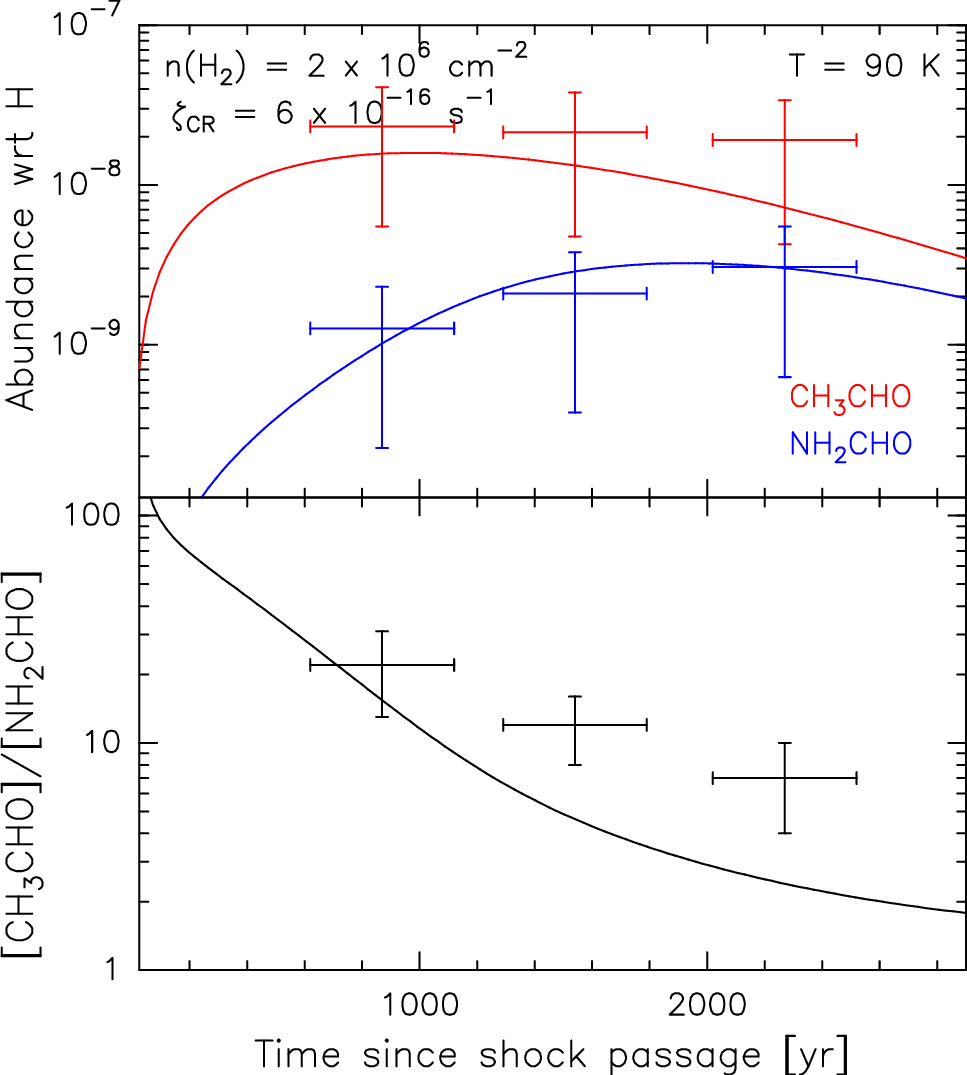}\\
    \includegraphics[scale=0.42]{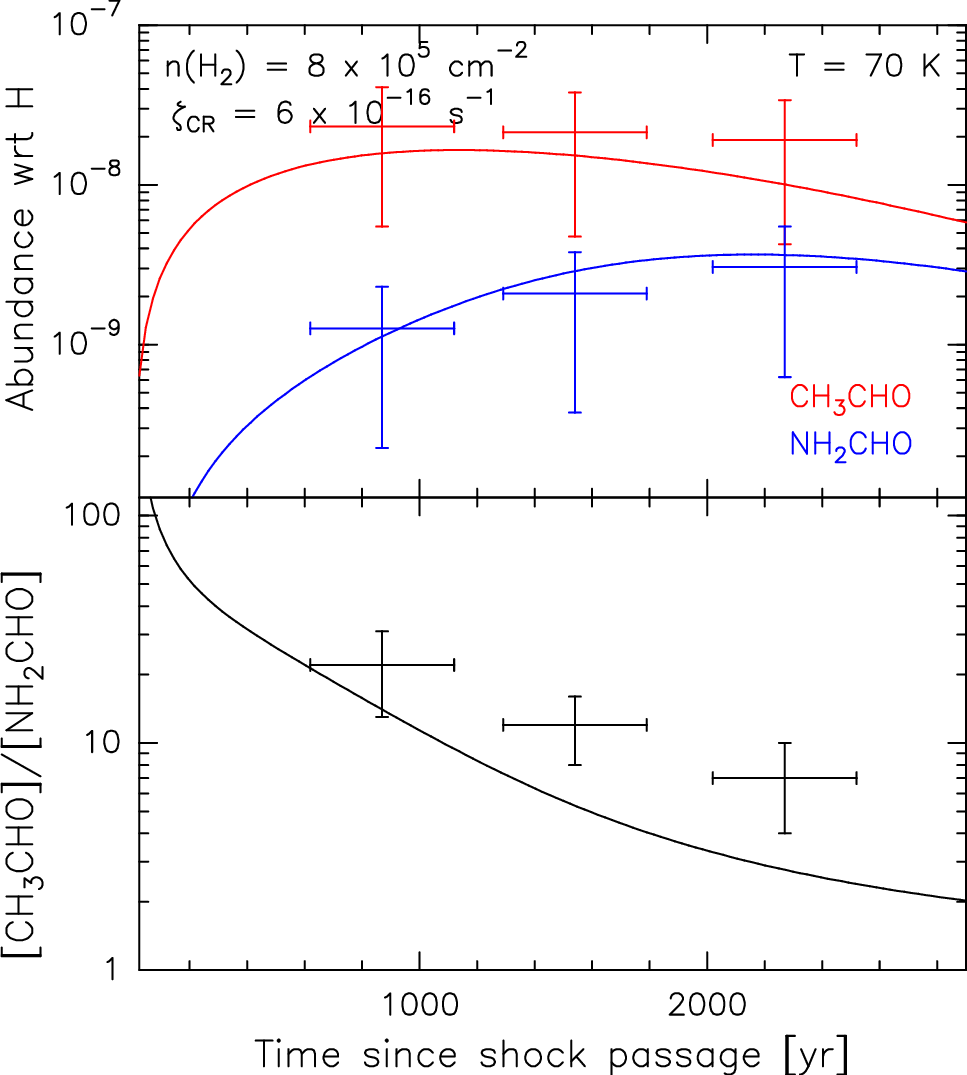} & & \includegraphics[scale=0.42]{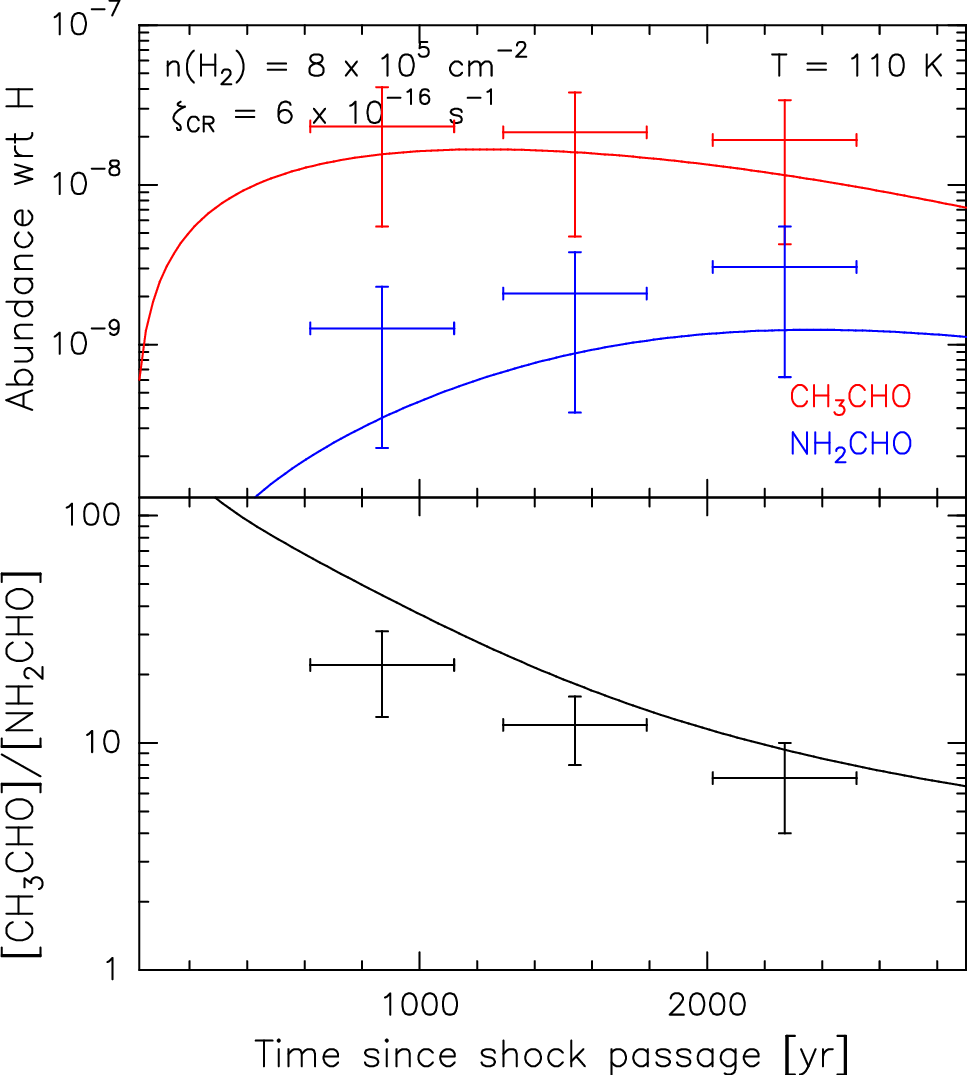}\\
    \includegraphics[scale=0.42]{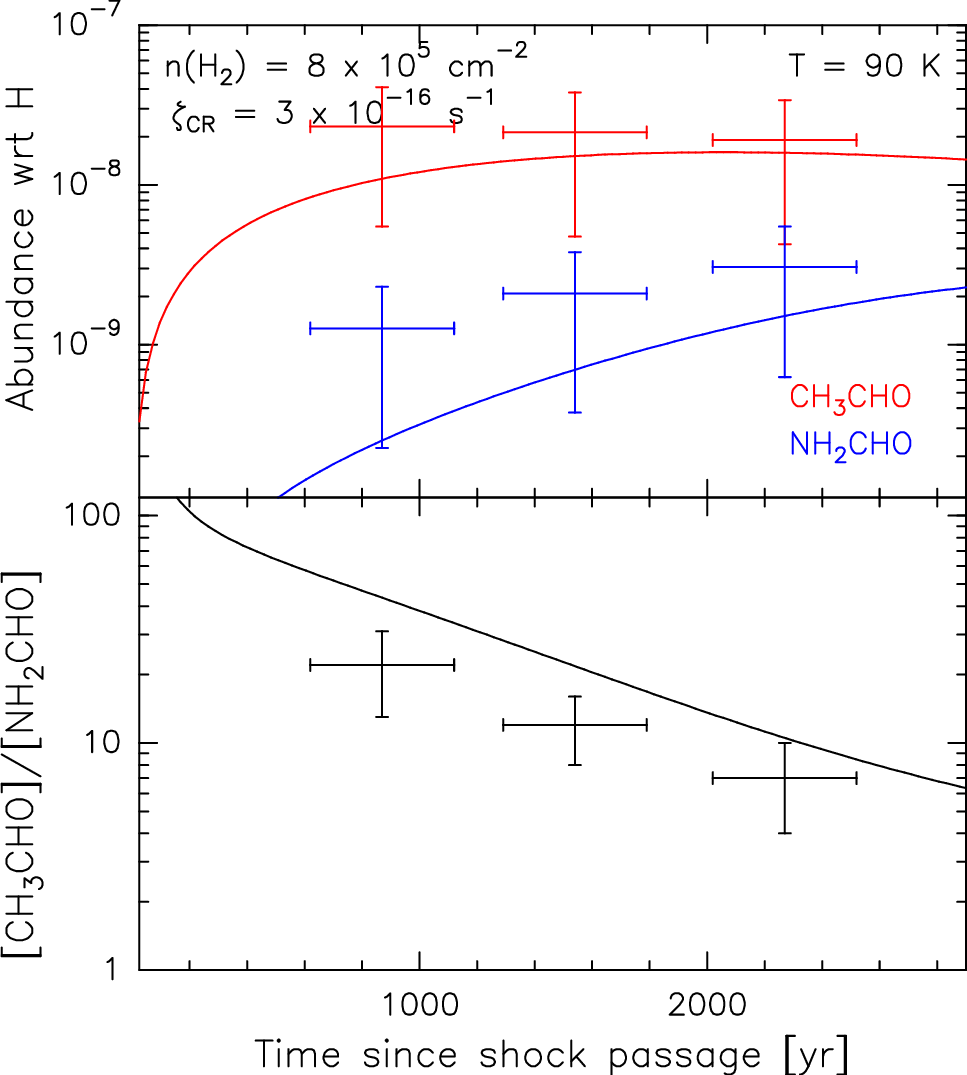} & & \includegraphics[scale=0.42]{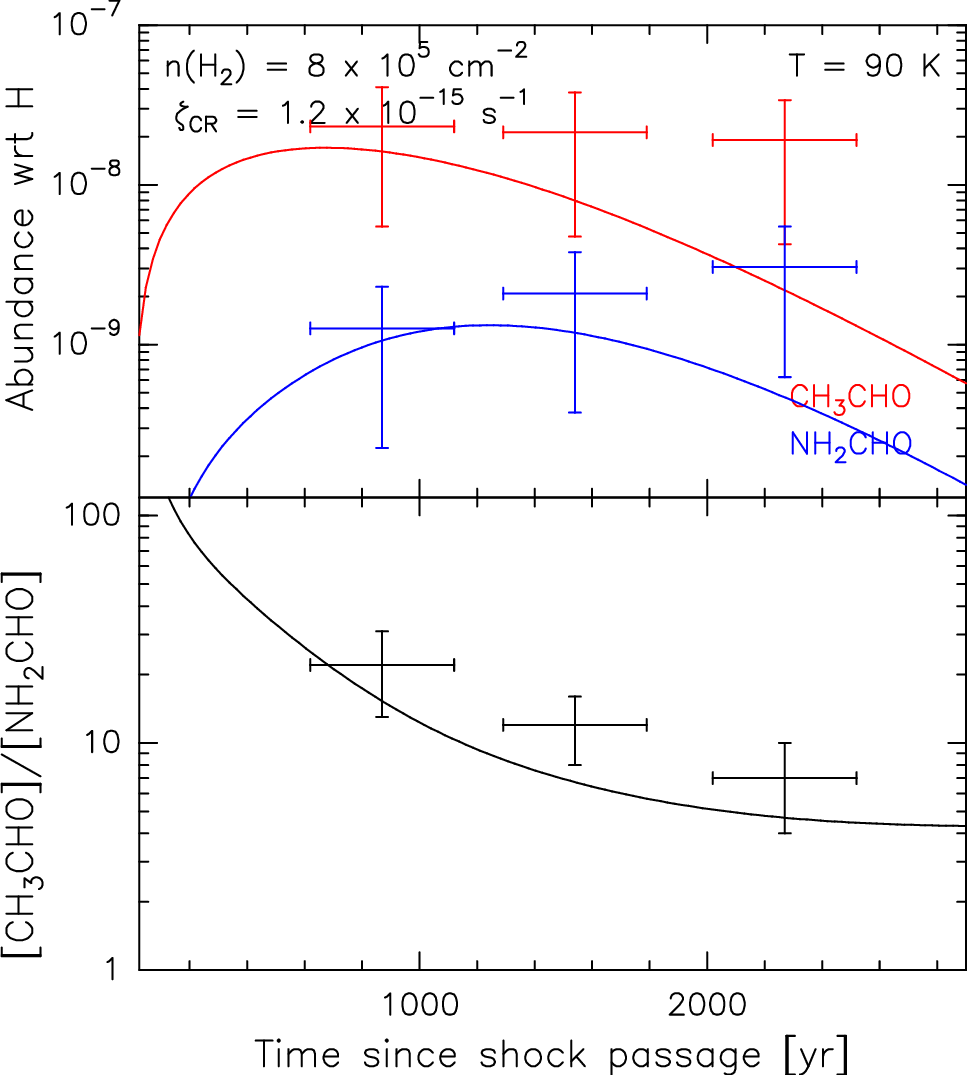}\\
   \end{tabular}
   \caption{Each double-panel box plots the CH$_3$CHO and NH$_2$CHO abundances (\textit{top panel}) and their ratio (\textit{bottom panel}) predicted by the gas-phase chemistry model for varying physical parameter values, and for injected molecular abundances equal to those that best represent our data (see Fig.\,\ref{fmodgas}). The top boxes vary the density, the central boxes vary the gas temperature, and the bottom boxes vary the CR ionisation rate. Measurement points in B0, B1, and B2 are shown as crosses whose vertical lengths correspond to the measured abundance (ratio) uncertainties (see Table\,\ref{tcoldens}), and whose horizontal lengths account for a 500-yr uncertainty in the post-shock time as derived by \cite{podio16}.}
              \label{fmodphys}%
    \end{figure*}

\end{appendix}

\end{document}